**Bulk-Switching Memristor-based Compute-In-Memory Module for Deep Neural Network Training**

*Yuting Wu, Qiwen Wang, Ziyu Wang, Xinxin Wang, Buvna Ayyagari, Siddarth Krishnan, Michael Chudzik and Wei D. Lu\**

Y. Wu, Q. Wang, Z. Wang, X. Wang, Prof. W. D. Lu

Department of Electrical Engineering and Computer Science, University of Michigan, Ann Arbor, MI 48109, U.S.

B. Ayyagari, S. Krishnan, M. Chudzik

Applied Materials, Inc., Santa Clara, CA 95054, U.S.

E-mail: wluee@umich.edu

The need for deep neural network (DNN) models with higher performance and better functionality leads to the proliferation of very large models. Model training, however, requires intensive computation time and energy. Memristor-based compute-in-memory (CIM) modules can perform vector-matrix multiplication (VMM) in situ and in parallel, and have shown great promises in DNN inference applications. However, CIM-based model training faces challenges due non-linear weight updates, device variations, and low-precision in analog computing circuits. In this work, we experimentally implement a mixed-precision training scheme to mitigate these effects using a bulk-switching memristor CIM module. Low-precision CIM modules are used to accelerate the expensive VMM operations, with high precision weight updates accumulated in digital units. Memristor devices are only changed when the accumulated weight update value exceeds a pre-defined threshold. The proposed scheme is implemented with a system on chip (SoC) of fully integrated analog CIM modules and digital sub-systems, showing fast convergence of LeNet training to 97.73%. The efficacy of training larger models is evaluated using realistic hardware parameters and shows that that analog CIM modules can enable efficient mix-precision DNN training with accuracy comparable to full-precision software trained models.

Additionally, models trained on chip are inherently robust to hardware variations, allowing direct mapping to CIM inference chips without additional re-training.

**1. Introduction**

Artificial Intelligence (AI), particularly deep neural networks (DNNs), has promoted advances in broad fields from language translation and dialogue[1] to image processing[2] and even protein structure prediction[3]. In general, DNN model performance improves by training bigger networks on larger datasets. However, training larger DNN models comes at an enormous cost and generates a large carbon footprint. While progress has been made to accelerate inference tasks through network compression[4] and hardware acceleration[5], the process of DNN training has remained largely unchanged. Vast amount of data needs to be shuttled back and forth between memories and processors in the training process, which makes the training process slow and power intensive.

Accelerating DNN training will require addressing the memory bottleneck of the conventional von Neumann architecture. To this end, memristor-based compute-in-memory (CIM) systems are promising solutions. The network weights can be stored as the analog conductance states of the devices, as a result of internal ion (e.g. metal ions[6], oxygen vacancy $V_{OS}$[7]) re-distribution in the switching layer under the external electric field. VMM, the core operation of DNN workloads, can be performed with memristor crossbars in situ in a single time step[5c, 8]. Many prior studies have demonstrated[5a, 5b, 5d, 9] that DNN accelerators built with high-density memristor crossbar arrays can achieve high performance and high energy efficiency for inference tasks. However, studies have shown that although weights can be reduced to ternary or even 1 bit precision[10] with acceptable accuracy during inference, weight updates during the training process need to be computed in floating point 32bit (FP32) precision[11] to achieve desired model accuracy. When using CIM systems to perform DNN model training, the limited granularity of device conductance changes, along with other non-idealities such as nonlinear and asymmetric conductance updates, will lead to significant accuracy loss or failure to converge. Other properties including low operating current, low noise, high endurance, reasonable retention, are also desired.

Multi-Level Cell (MLC) capability has been shown by filamentary resistive RAM (f-RRAM)[6d, 12] and phase-change memory (PCM)[13], but predictable analog switching behavior is challenging due to the abrupt, stochastic filament formation process[7b, 14]. Electrochemical organic devices offers linear and symmetric conductance update with >500 conductance levels[15], but array configuration and system-level integration are still lacking. Recently, we reported a new class of forming-free bulk RRAM (b-RRAM) featuring up to 128 programmable levels with high uniformity and yield across 300mm wafers[16]. The device operates by modulating the profile of oxygen vacancy concentration in the bulk switching layer, therefore enabling gradual and precise programming. A System-on-Chip (SoC) design that integrates 4 tiles of b-RRAM crossbar arrays together with essential peripheral circuitry and a RISC-V processor[9a] have also been demonstrated for inference applications[16].

In this work, we apply the memristor-based CIM modules for DNN training tasks, using a recently proposed mixed-precision scheme[17]. The approach, based on a hybrid of CIM modules and digital systems, is shown in **Figure 1**, where the analog CIM performs VMM operations during forward propagation, and the gradient update during backpropagation is computed and accumulated in the high precision digital system. The memristor conductance will only be updated when the accumulated gradient is larger than the device programming granularity, therefore greatly relaxing the hardware requirements for conductance update precision and write endurance. Recent studies have demonstrated this mixed-precision training scheme with phase-change memory[17-18], and shown that reasonable accuracies can be obtained even with noisy and low-precision weight update. In this study, the employment of fully integrated CIM modules with b-RRAM devices that have high uniformity and fine-grained weight updates ensures the conductance updates can be reliably obtained and lead to fast convergence during training.

Specifically, we train a LeNet model on chip and show that 97.73% accuracy can be achieved for MNIST classification in only 13 training epochs. To evaluate the system performance for larger models, we develop an accurate hardware-aware AI simulator that considers all levels of hardware constraints. After incorporating realistic device statistics, the training evolution of the LeNet model predicted by the AI simulator matches well with the experimental results. We then show that the proposed mixed-precision training scheme using realistic CIM systems can approach iso-accuracy of pure digital software training, for models such as VGG-8 and ResNet-18 for the CIFAR-10 dataset. Additionally, models trained using the mixed-precision method are inherently robust to hardware non-idealities, allowing direct transfer of the trained weights to other CIM chips without additional model retraining stage. These results suggest the b-RRAM CIM module is a promising candidate for efficient neural network training acceleration using the mixed-precision scheme.

## 2. Experiment and Results
### 2.1. Analog-Digital Heterogeneous System

The system is schematically illustrated in Figure 1. It is composed of analog memristor CIM modules for efficient VMM operation and a high-precision digital unit for gradient computation and weight update accumulation.

To map the network model on chip, the convolution kernels are unrolled into 1D vectors and converted to device conductances. Since the conductance can only be positive, each weight is split into positive and negative parts and differentially programmed to a pair of devices on adjacent columns (using the dual-column scheme). In the forward pass, the input vectors are applied to the drive lines as voltage pulses. Following Ohm's law and Kirchhoff's current law, the current at each column represents the dot product of the input voltage vector and device conductance vector $I_j = \sum_i x_i w_{i,j}$. The current output of the negative column is subtracted from that of the positive column to produce the final VMM output. The output currents are then converted to voltages and digitized. The digital outputs $z^l$ are sent back to the digital unit for intermediate operations (activations, pooling, batch-norm) and applied as inputs to the crossbar array in the next layer $x^{l+1}$.

Before the training process, the initial device conductances are read out and stored in the digital unit. The high precision weight copy, $W_{FP}$, is later used to compute the gradient in the backward pass. After the forward pass is completed through the CIM module, we obtain the loss function by comparing the network output to the ground truth. The error and gradients are calculated following the standard chain rule in the backward pass, as $\delta^l = (W_{FP}^l \delta^{l+1}) \odot \sigma'(z^l)$; $\Delta W_{new}^l = \eta x^l \delta^{l+1}$, where $\delta^l$ is the error term, $W_{FP}^l$ is the high-precision weight matrix, $\Delta W_{new}^l$ is the high-precision weight update of the current batch, $x^l$ is the input, $z^l$ is the VMM output of layer $l$, $\sigma$ is the activation function and $\eta$ is the learning rate. The weight update is calculated after each batch and accumulated in the high precision weight change matrix $\Delta W_{FP}^l$ in the digital system, as $\Delta W_{FP}^l \mathrel{+}= \Delta W_{new}^l$.

When the accumulated value in $\Delta W_{FP}^l$ is over a pre-defined threshold, the corresponding memristor device will be updated through a write-and-verify scheme to the new target value $w_{FP}$ $+ \Delta w_{FP}$, and the corresponding value in $\Delta W_{FP}^l$ will be reset to zero. To reduce training time, we define the maximum Set-Reset trial cycle to 2 during RRAM update. In other words, even if the device does not fall into the desired range after the maximum number of trials, we will not further fine tune the conductance. After the weight update, the corresponding element in the digital weight matrix $W_{FP}$ will be updated to the new theoretical weight $w_{FP} + \Delta w_{FP}$, while the other elements remain the same. Note that the digital weight $W_{FP}$ may be different from the actual device conductance $W_{RRAM}$ due to program error and conductance drift, but this deviation is small due to the precise programming ability and the stability of the b-RRAM devices, as shown in **Figure 2**. Any small deviations are also reflected and accounted for during the mixed-precision training algorithm.

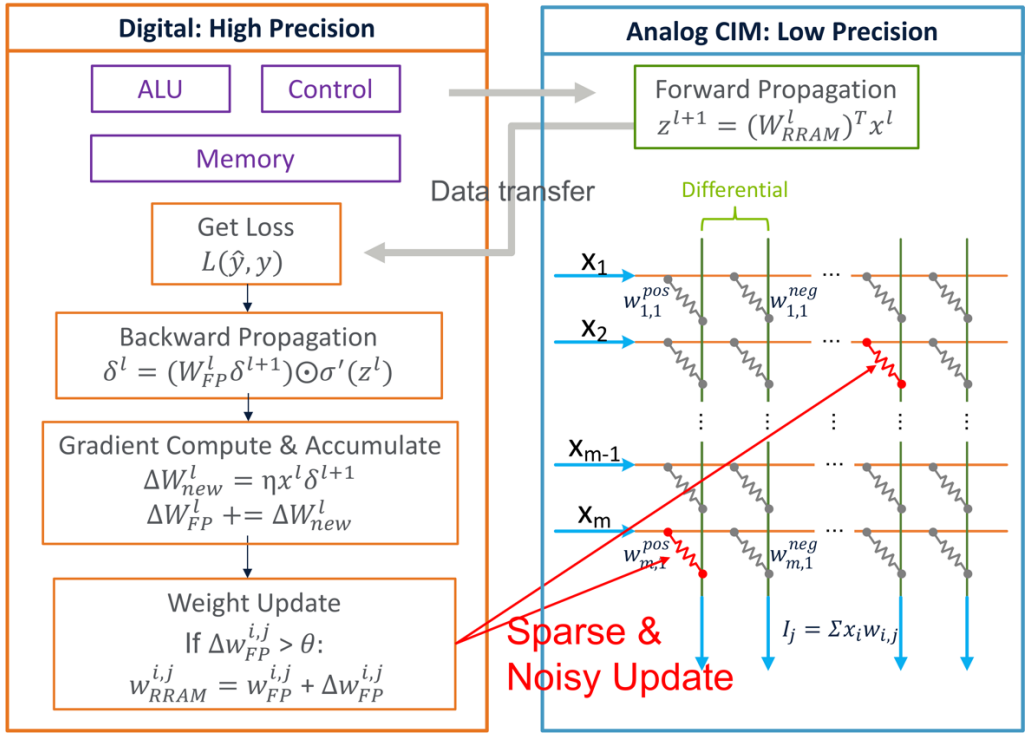

**Figure 1.** Schematic overview of the hybrid CIM/digital system for deep neural network training. Weights of the neural networks are differentially mapped to memristor conductance ($w_{i,j}^{pos} - w_{i,j}^{neg}$). The VMM operation in the forward pass is performed with the CIM module. The results are sent to the digital unit to compute the gradient in the backward pass. The weight updates of layer l are accumulated in $\Delta W_{FP}^l$. If $\Delta W_{FP}^l$ exceeds a defined threshold, the corresponding memristor device will be programmed to the new target conductance level, and the element in $\Delta W_{FP}^l$ will be reset to zero.

## 2.2. Analog Bulk RRAM

The proposed training scheme still requires reliable multi-level resistance switching. Switching of filamentary RRAM (f-RRAM) is typically abrupt and stochastic, since the conductance change is dominated by the movement of only a few ions in the conductive filaments[6a, 7b, 14b]. In a recent work[16], we introduced a non-filamentary bulk RRAM (b-RRAM) device, and showed that b-RRAMs can be reliably programmed up to 128 levels between 400nA and 4μA (4-40μS) across 300mm wafer[16] (Figure S1).

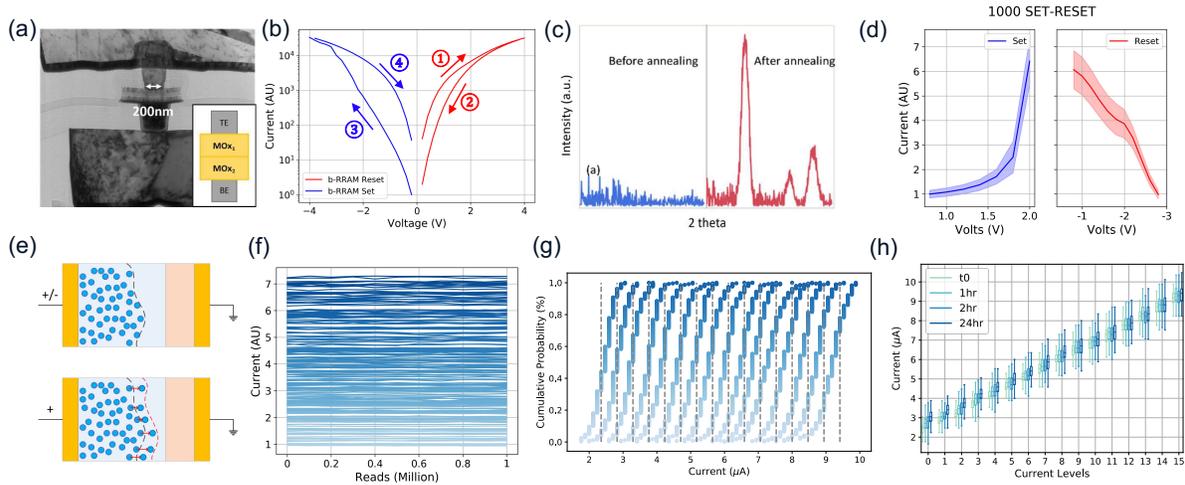

**Figure 2.** (a) TEM image of a b-RRAM stack (device diameter, 200nm) integrated between the M4 and M5 layers in the BEOL process. Inset shows the bilayer RRAM stack. (b) DC I-V curve of a typical b-RRAM. Arrows show the sweep sequence. The as-fabricated b-RRAM devices are in the on state. (c) XRD data of the metal oxide film: for as deposited and post-annealed states. (d) DC cycling of the device between LRS and HRS for 1,000 cycles. The envelope represents the ensemble of the currents, and the solid line is the average of the 1,000 cycles. (e) Schematic of the bulk switching mechanism. The device operates by modulating the oxygen vacancy concentration profile in the bulk volume. (f) Device stability through 1,000,000 read cycles. (g) CDF plot of 16 levels programmed on the 4,096 devices in one tile. (h) Boxplot showing the distribution of the 16 programmed levels at room temperature over 24 hours.

The b-RRAM cells in this study are fabricated in a 16kb 1T1R array with 65nm CMOS technology. **Figure 2**a shows a transmission electron microscope (TEM) image of the b-RRAM stack, integrated between the M4 and M5 layers in a standard back end of line (BEOL) process using a custom 300mm deposition system. The deposition system is specially built by Applied Materials to enable sequential physical vapor deposition and atomic layer deposition without air-break. The fabrication process is carefully controlled to ensure high uniformity and yield across the 300mm wafer.

The b-RRAM device is composed of a bilayer stack of wide-band dielectric and metal oxide, respectively. The metal oxide serves as the reservoir of oxygen vacancies ($V_{OS}$), and the wide-band dielectric functions as a tunneling barrier to control the current level. Following the film deposition, the wafer goes through post-deposition annealing treatment to induce the phase transition of the metal oxide film to a certain crystal phase as shown by the X-ray diffraction analysis (XRD) in Figure 2c. This particular phase has inherently high $V_O$ concentration, which is critical for forming-free and bulk type switching[16]. More details on the device fabrication and material characterization can be found in [16].

The device after fabrication is in the on state. We first scanned the device in the Reset direction, then followed with the Set direction. The DC I-V curve in Figure 2b shows the switching is smooth and gradual. We further tested the device cycle-cycle reliability by DC cycling the device between the low resistance state (LRS) and the high resistance state (HRS). Figure 2d shows that the device can be reliably switched with good cycle-to-cycle uniformity as desired for the mixed precision training process. These results support the bulk switching mechanism, where resistance switching of the b-RRAM is governed by the profile of defects ($V_{OS}$) in the bulk of the reservoir layer instead of the discrete, localized defects in conductive filaments, as shown in Figure 2e.

We programmed the devices to different current levels and read continuously for 1 million cycles (Figure 2f). All conductance levels stay stable, exhibiting low noise level and robustness to read disturb. The low noise is enabled by the bulk switching characteristics, where the random motions of individual $V_O$ are essentially averaged out. Figure 2g shows the distribution of 16 levels (2.3-9.4μA/7.6-31.3μS at 0.3V) from 4,096 devices in one array using an incremental step pulse programming scheme. The flow chart of the programming scheme is depicted in Figure S2. Since the devices are born in the on state, we first reset all devices to the lowest conductance level. Then adaptive set pulse train is applied to the devices under test from 1.5 to 3V. After each pulse, we read out the conductance value to check whether it reaches the set threshold. If so, we jump out of the set process and switch to the adaptive reset sequence. The reset voltage pulses start from 1.5V to 3V to fine tune the device conductance. If the device falls into the desired target range, the programming process is completed. If the device is over reset, we restart from the set process. The maximum cycle number of Set-Reset trials is set to be 5 in this endurance test. Later in the on-chip training process, we reduced the maximum Set-Reset trials to 2 to minimize the training time and energy consumption. The cumulative distribution function plot (CDF) in Figure 2g shows the tight distribution of each level and 100% array level yield. Figure 2h plots the retention characteristics for the 16 different resistance levels measured at room temperature. The medians in the low current range shift up slightly after 2hrs, but the distributions remain tight even after 24hr, which is sufficient for the proposed training applications.

### 2.3. Hardware Testing System

The b-RRAM crossbars are heterogeneously integrated with the CMOS circuit underneath. The CMOS circuit, a system-on-chip (SoC) in this case, is designed to enable full control of the crossbar array and minimize delay during signal transmission. The SoC includes essential mixed-signal circuitry to form a complete CIM module with the b-RRAM crossbar, routing circuitry that integrates 4 tiles of RRAM CIM modules, a RISC-V processor with 32kb instruction memory and 512kb data memory, and a phase-locked loop (PLL) to generate a high-frequency clock for the SoC. A microscope photo of the SoC die is shown in **Figure 3**a.

Each tile is an independent self-contained module including a 64×64 1T1R b-RRAM array and mixed-signal circuitry (Figure 3c). 64 digital-to-analog converters (DACs) are used to quantize the inputs to 8-bit and apply them to the drive lines (DLs) of the CIM module in parallel, and a 10-bit word line (WL) DAC is used to control the gate voltage of the access transistor in the 1T1R b-RRAM cell. The bit line (BL) of each column is connected to a transimpedance amplifier (TIA) and an 8-bit analog-digital-converter (ADC). The sense-amplifier circuit is shared by a group of 8 BLs, so 8 operations are needed to read out the multiply-accumulate (MAC) results of the entire array. The DL, BL and WL switches together support set/reset/read/MAC operations, controlled by the host processor via fast direct memory access (DMA). To further boost the CIM throughput, a local timing controller synchronizes the analog compute independently from the host processor. The modular design can be scaled by adding more tiles through the AXI interface, and different RRAM array sizes can be designed to accommodate different DNN models.

One key feature of the circuit design is the bit-serial operation assisted by the binary-weighted multi-cycle sampling ADC. In the bitwise approach, eight 1-bit pulses of constant voltage are applied by the DL DACs, where the $n^{th}$ pulse corresponds to the $n^{th}$ bit of the input. The ADC weights the partial bitwise outputs in every cycle and digitizes the weighted sum in one conversion step. 9 cycles in total are needed to produce a single 8bit output. For more details of the circuit design and pulse scheme, please check our previous works[5d, 9a].

The testing setup includes a custom printed circuit board and a Raspberry Pi microprocessor to provide the voltage and current references, digital control and data transfer functions (Figure 3b). The Raspberry Pi can access the CIM SoC through IO peripherals including serial peripheral interface (SPI), universal asynchronous receiver transmitter (UART) and general-purpose input/output (GPIO). The communication between Raspberry Pi and the host computer is established through the USB or ethernet connection. A SPI flash is also included on board for dataset storage.

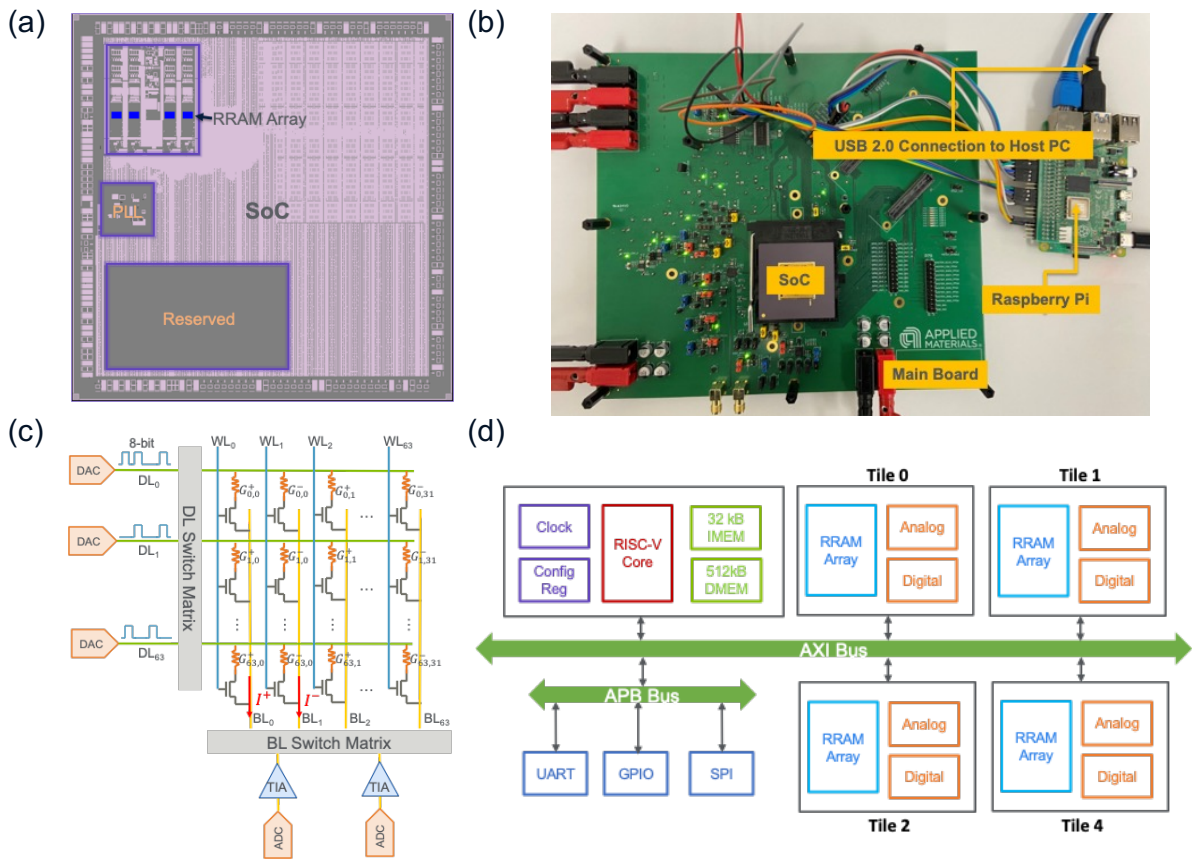
9

**Figure 3.** (a) Die photo of the CIM SoC. Four RRAM arrays are integrated on one chip. (b) The custom designed PCB for chip testing and DNN training applications. A Raspberry Pi is used for the CIM SoC control and communication between the CIM SoC and the host PC. (c) Schematic of the RRAM CIM tile. The 64×64 1T1R array is connected to 64 8-bit row DACs. One TIA and one 8-bit ADC are shared by 8 columns. (d) Block diagram of the full system. Four CIM tiles are attached to the AXI bus as self-contained IPs. An on-chip RISC-V core controls the programming and computing of the CIM modules.

A hardware simulator is developed in PyTorch to analyze the experimental mixed precision training results and simulate the training process of larger DNN models. Different hardware constraints must be carefully considered to mitigate the accuracy degradation in DNN training. In a previous work, we proposed an architecture-aware model optimization scheme for DNN inference evaluation that organized the hardware limitations into three levels[5c, 19] as illustrated in **Figure S4**. In this paper, we adopted the similar hierarchical organization and enabled DNN training from scratch. Level-1 only incorporates quantization effects and device non-idealities including read noise and limited on/off ratio. To support quantization-aware training, a fake-quantization function is defined in PyTorch that takes in high precision weights and outputs quantized values. The gradients are computed with the commonly used straight-through estimator[20]. Weight noises are injected as random samples from a defined Gaussian distribution and added to the weight matrix in every convolution operation. In Level-2, the dual column mapping of positive and negative weights is included. Level-3 considers the effects of finite array size, as shown in Figure 4b. Since the crossbar array has limited number of devices, the kernels in the DNN models are generally too large to fit on a single crossbar. The practical and scalable approach is to split the weights into multiple crossbar tiles. The inputs are also sliced and applied to each crossbar. The output activations are computed by summing up the partial sums from each array. The outputs will then be multiplied by a scaling factor, converted to output activations and applied to the crossbar arrays in the following layer as inputs. In architecture-aware training, the scaling factor at each crossbar is a trainable parameter. By including these three levels of non-idealities in our training platform, we can essentially mimic the convolution operations on chip and make realistic evaluation of training performance.

The simulator implements the mixed-precision training processes the same way as the hardware set up. It stores two weight matrices—one is digital weight matrix W$_{FP}$ and the other is the accumulated weight change matrix ΔW$_{FP}$. The gradient is computed in FP32 precision based on the on-chip activations and digital weight W$_{FP}$. The calculated weight updates of each batch are accumulated in the weight change matrix ΔW$_{FP}$. When an element in ΔW$_{FP}$ exceeds the predefined threshold, the corresponding RRAM device value will be updated. During weight update, we add a programming error based on the device statistical model to account for the stochastic effects and cycle-cycle variations.

In the following session, we evaluate the efficacy of the hybrid system via experimental demonstration of LeNet training and hardware-aware simulation of VGG-8 and ResNet-18. For on-chip testing, the grey rectangular part (forward propagation) in Figure 4b are performed on chip.

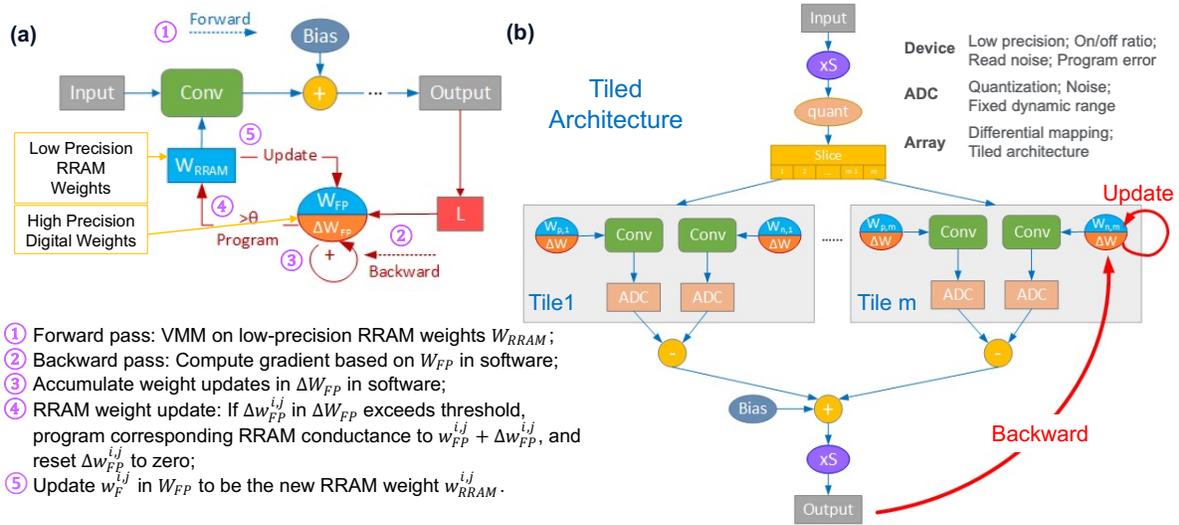

**Figure 4.** Hardware simulator for mixed-precision DNN training. (a) Simplified block diagram of the overall training approach, neglecting details of hardware constraints. In the forward pass, a batch of inputs are sent to the analog RRAM arrays for VMM operations. In the backward pass, the gradients are computed based on the high-precision weight copy in the digital unit. The weight updates are accumulated and propagated to the corresponding RRAM devices if they are over a defined threshold. (b) Detailed block diagram showing three levels of realistic physical constraints for practical DNN training on memristor-based CIM. Level-1: device non-idealities and quantization effects. Level-2: signed weight representation in dual column. Level-3: limited

array size and ADC effects (fixed current range and noise). A custom optimizer is defined to support weight update accumulation. Programming noise is injected during the conductance update.

## 2.5. On-chip Training of LeNet

To verify the efficacy of the proposed hardware-software training approach, we experimentally demonstrate the training of LeNet, a three-layer convolutional neural network (CNN) for handwritten digit classification. The network architecture is depicted in **Figure 5**a. For weight update, we assume a conservative programming granularity of 2bit with 4× on/off ratio (memory window) between 0.82-3.29μA (8.2-32.9μS). In each training epoch, we select 400 randomly shuffled batches of images, and each batch contains 64 images. The device initial conductance lies within 0.82-2μA (8.2-20μS). All initial conductance values are read out, scaled to the network weights and stored in the digital unit for the backward propagation. For convolution operations, a 5×5 sliding window moves across the input image, and the windowed region is flattened into a 1D vector and quantized by the 8-bit DACs on the DLs. The RRAM CIM performs the VMM operation, and the activation outputs are collected and sent back to the Raspberry Pi through SPI. The digital circuit will then perform the intermediate processing, including the bias addition, multiplication of the scaling factor and activation functions.

After the forward pass is completed, we start the gradient computation in high precision in the digital unit. The loss function is computed between the on-chip results and the ground truth label. Starting from the output layer, the gradients of the loss with respect to the network parameters are computed following the backward propagation, and weight updates are calculated with the defined optimizer and accumulated in $\Delta W_{FP}$. Here, we use Adam [21] with weight decay as the optimization algorithm, and the initial rate is set as 0.004. When the magnitude of $\Delta w_{FP}$ is increased over the RRAM programming granularity, we transfer the weight update to the corresponding RRAM device following the incremental step pulse scheme described earlier. The maximum Set-Reset cycle is defined as two trial cycles. The element in the digital weight will be updated to the new ideal value $w_{FP} + \Delta w_{FP}$. The other elements whose corresponding devices are not updated will not be changed, even if they may deviate from the actual RRAM conductances due to programming error or retention. In this way, we do not need to read the entire array after each epoch, which is very time and energy consuming.

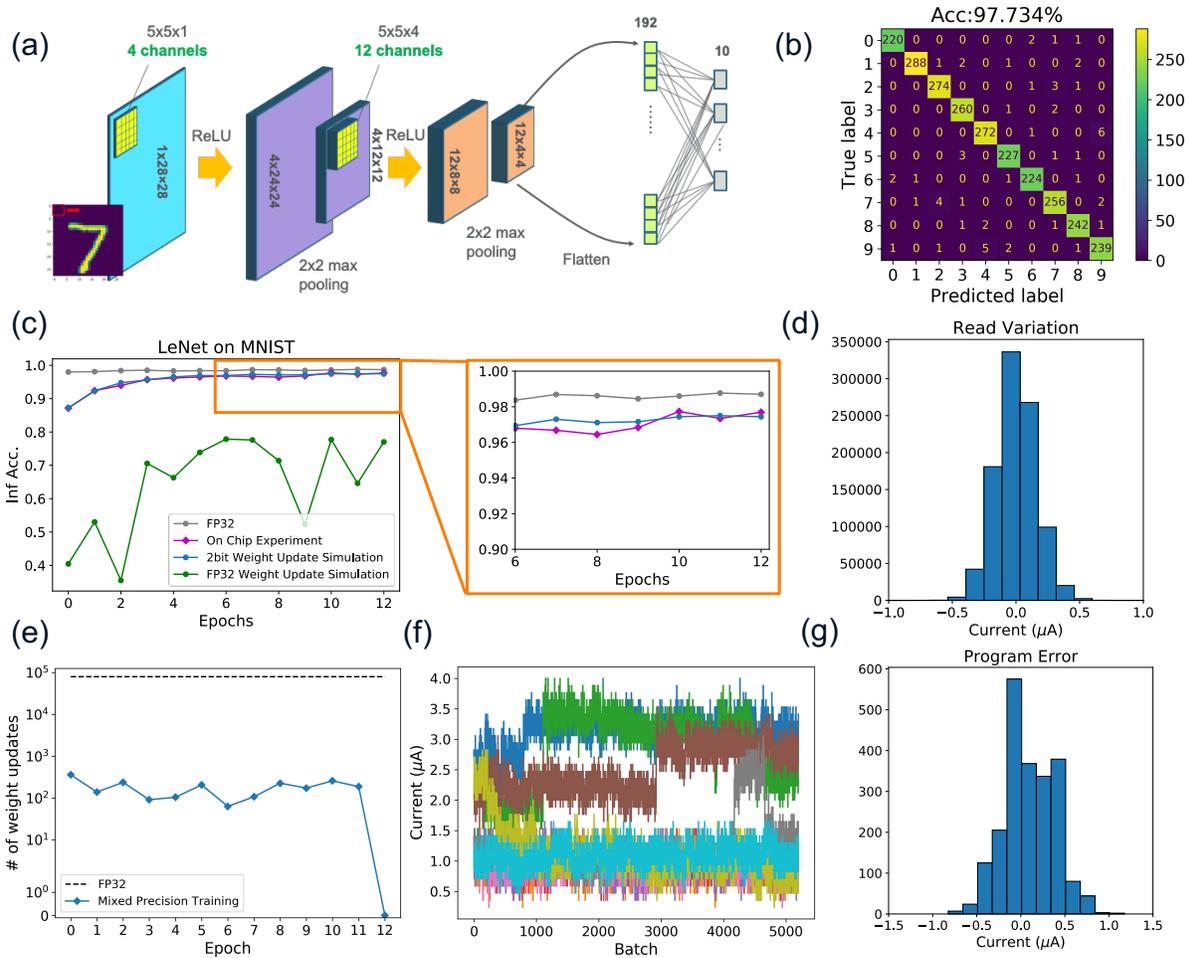

**Figure 5.** (a) Schematic showing the LeNet architecture for MNIST classification. It consists of two convolution layers and one fully connected layer. (b) Confusion matrix with the network trained on chip tested over 2560 image samples. (c) Left: evolution of the inference accuracy over 13 training epochs. Right: Zoom-in figure of the marked region. Gray: Software training results with FP32 precision, Magenta: on-chip training results, Blue: Simulated RRAM training results with realistic hardware characteristics, Green: Simulated RRAM training results with software training approach. (d) Histogram of read variation. (e) The total number of weight updates per training epoch. (f) The evolution of 10 weights over the training process. (g) Histogram of the weight program error.

Figure 5f shows the weight evolution of 10 randomly selected weights, showing that the weight updates are sparse. 0.1V is selected as the read voltage during weight programming and VMM operations to reduce the power consumption. The readout of the device conductance is noisy, as reflected by the fluctuation in the weight evolution plot. This is mainly due to the limited ADC resolution instead of the intrinsic device noise, as validated in Figure 2f. The histogram of read variation is shown in Figure 5d, which contains device read variation, conductance drift, ADC quantization error and ADC noise. Thanks to the gradient accumulation method, only a few weights need to be updated after each batch, whereas every weight needs to be updated in the software training (FP32). As is shown in Figure 5e, the number of weight updates are dramatically reduced by around 500×, and each weight only needs to be updated 10.8 times on average. The weight programming error is plotted in the histogram of Figure 5g. Despite the read variation, we can still achieve precise conductance update thanks to the bulk-switching characteristics.

Compared to the ideal software training (grey line), the on-chip training yields lower inference accuracy in the first few epochs due to the sparse update and low-precision weights. However, it quickly catches up after 6 epochs and converges after 13 epochs. Afterwards, the number of weight updates drops to zero, so we stop the training process. The inference accuracy is measured based on 2560 test images after each epoch (Figure 5b). Accuracy as high as 97.73% is achieved within 1% accuracy drop compared to the software model (98.7%).

To evaluate the efficacy of the hardware simulator, we feed the device and circuit statistics measured on chip to the simulator. As depicted in Figure 5c, the simulated training process (blue line) matches very well with the actual experiment results (magenta line). This validates that the hardware simulator can capture the complete set of hardware characteristics successfully. Therefore, realistic prediction of neural network performance on chip can be performed. We also simulate the case when we train the network with b-RRAM devices using the naïve software training approach without gradient accumulation (green line). It is not surprising to see the model fails to converge, as the weight update at each batch is too small to be reliable represented by the RRAM conductance. The highest achievable accuracy is only 77.8%.

We show that the hardware-software combined training approach makes it feasible to train neural network using b-RRAM CIM and achieve accuracy comparable to software baseline even with limited weight precision and the intrinsic noise of analog computing system.

Additionally, it significantly relaxes the requirement for endurance and reduces the weight update overhead, therefore accelerating the training process with improved power efficiency.

### 2.6. Simulation of Deep Neural Network Training

To evaluate the feasibility and efficacy of training with b-RRAM CIM for deeper networks, we perform analysis using the hardware-realistic simulator. All three levels of hardware constraints are considered in the simulation with realistic hardware settings. The assumptions are shown in the **Table 1**, where the parameters are defined based on experimental measurements. The b-RRAM precision is set to be 16 levels with on/off ratio as 7, which has been demonstrated on chip as shown in Figure S3. The update threshold is set as 1/15 of the RRAM conductance range, corresponding to the 4-bit weight precision. The standard deviation of write error, read variation and ADC noise are extracted from on chip testing results. Two benchmark models are used for evaluation: an 8 block VGG and ResNet-18 for CIFAR-10 classification. The network architectures are depicted in **Figure 6**a and f.

To accommodate larger networks and improve power efficiency, a larger crossbar array size of 256×64 is used in the simulation. Further increase of the crossbar array size may not be feasible, as the I-R drop increases with the number of rows, leading to inaccurate voltage delivery and current accumulation. The convolution weight matrix will be mapped to tiled arrays in the dual column configuration.

**Table 1.** Parameter settings used for the hardware-aware simulation of DNN training

| Crossbar size | ADC bits | ADC current range | Std of ADC noise | RRAM bits | RRAM current range | Std of RRAM read variation | Std of RRAM program error |
|---|---|---|---|---|---|---|---|
| 256x64 | 8 | 0-70μA | 2σ[a] | 4 | 1-7μA | 0.3σ | 0.5σ |

[a] σ is defined as the separation between two adjacent levels. For example, 2σ means the std of ADC noise equals $2 \times 70\mu A / (2^8 - 1) = 0.549\mu A$.

Adam with weight decay[21] is used as the optimizer. The initial learning rate is set as 0.003 and the learning rate is reduced by half if the test accuracy has stopped improving for consecutive 5 epochs. 64 images are used for one training batch. The models are trained for 100 epochs. Figure 6b compares the accuracy of the mixed precision training approach to the software baseline. To account for weight initiation effects, 10 runs are performed from random initialization for both cases. The VGG-8 achieves software comparable test accuracy (0.9% accuracy drop), despite the fact that the initial convergence of the model is slower than the software model.). For ResNet-18, the evolution of the mixed precision training is similar to that of the software training. The accuracy after 100 epochs matches the software baseline (0% accuracy drop). In both cases, the models are able to converge even with 4bit RRAM programming resolution and achieve adequate accuracy in the presence of various hardware limitations.

The mixed precision training approach significantly reduces the number of write/erase operations on the RRAM devices. Figure 6d and h show that the number of weight updates occurred during the training process can be reduced by approximately 1000× when compared with the naïve implementation. As the training process proceeds, the learning rate decreases, which further lowers the number of weight updates per epoch. On average, the weights in every model only need to be updated less than a hundred times, and the highest programming times of one weight are less than 220 in the entire training process.

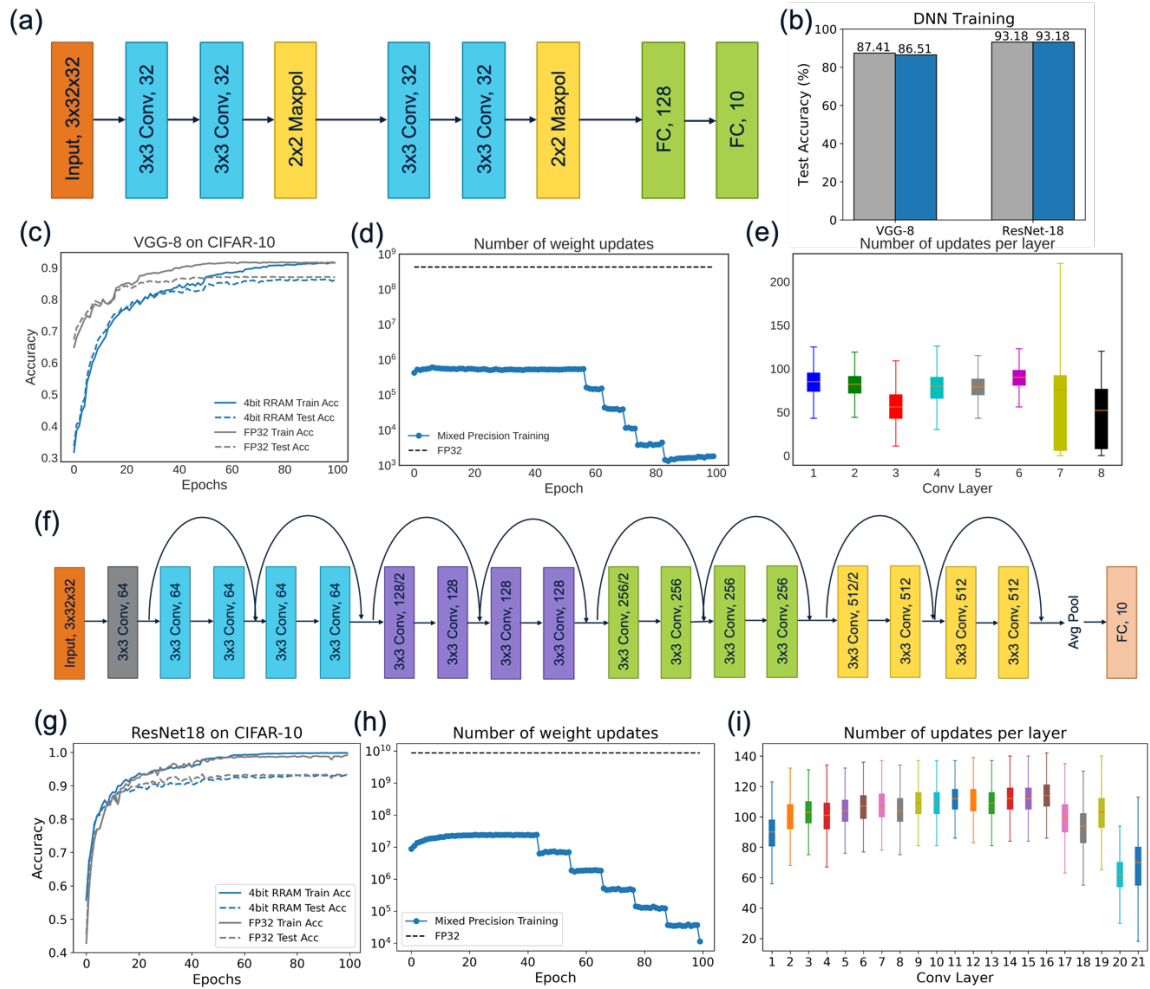

**Figure 6.** (a) Schematic showing the VGG-8 architecture for CIFAR-10 classification, composed of six convolution layers and two fully connected layers. (b) Accuracy comparison between the models trained with the mixed precision approach and software baseline. Realistic hardware imperfections are considered in the simulation. The proposed system achieves software comparable accuracy: 86.51% accuracy (0.9% accuracy drop) for VGG-8 and 93.18% accuracy (0% accuracy drop) for ResNet-18. Average of 10 models trained from random initialization is plotted. (c-e) VGG-8 training performance. (c) The evolution of the training accuracy and the test accuracy during training of VGG-8 for 100 epochs. Due to the hardware constraints, the accuracy corresponding to the RRAM-based training approach is lower than the software approach, but the model is able to converge to comparable accuracy after 100 epochs. (d) The total number of weight updates across all layers of VGG-8 in one training epoch. The gradient accumulation method reduces the number by about 1000× compared to the software training. (e)

Boxplot showing the distribution of the weight programming times in each layer of VGG-8 model. (f-i) ResNet-18 training performance. (f) Schematic showing the ResNet-18 architecture for CIFAR-10 classification. (g) The evolution of the training accuracy and the test accuracy during training of ResNet-18 for 100 epochs. (h) The total number of weight updates across all layers of ResNet-18 in one training epoch. The gradient accumulation method reduces the number by about 1000× compared to the software training. (i) Boxplot showing the distribution of the weight programming times in each layer of ResNet-18 model.

### 2.6. Evaluation of Transferred Model Accuracy

Although RRAM based CIM modules can perform VMM operations with high efficiency, the inference accuracy must be ensured before practical use becomes feasible. When a trained model is transferred to a new chip for NN inference, the mapped model will inevitably deviate from the original target model due to the weight programming error, leading to inference accuracy drop when compared to the originally trained model[13]. Unlike read noise, these weight deviations are static during inference. Here, we show on-chip training naturally include all hardware non-idealities, producing a network that is robust to weight deviation.

In **Figure 7**a, we evaluate how the models trained using the mixed-precision approach will perform after weight transfer, for LeNet, VGG-8 and ResNet-18. The accuracies are normalized based on the software baseline models. We perform 10 training trials from random initialization for each case. For the model trained in each trial, we sample random programming noise for 10 times with sigma = 0.5 (100 trials in total for each boxplot). The models trained with floating point precision suffer from significant accuracy drop after weight transfer. Quantization aware training in software improves the generalization of the model, but the accuracy drop is still very severe. The standard deviation between different quantized models is significant, meaning quantization aware training is sensitive to weight perturbation. In contrast, models trained with mixed precision training are inherently robust to weight programming variations, and can achieve software comparable accuracy even after weight transfer as shown in Figure 7b. We simulate the case with (magenta) and without (cyan) weight programming error when transferring the models to a new chip. The accuracy drop is negligible for LeNet and ResNet18 without any re-training. For VGG-8, less than 2% accuracy drop from original trained model is obtained.

To address weight transfer induced accuracy drop, retraining every inference chip (i.e. chip-in-the-loop retraining) was proposed. For smaller networks, it is sufficient to retrain only the last fully connected layer for each chip[22]. For larger networks, it was proposed to map the network progressively one layer at a time and fine tune the model after each layer is transferred[23]. Although this method can effectively recover the accuracy, the process is very time and energy consuming and needs to be performed for each individual chip, making the approach less practical. The mixed-precision training, on the other hand, incorporates all levels of hardware constraints including programming error and read noise during the initial network training process, making the resulting model tolerant to weight disturbance. Therefore, after training the model can be directly programmed to new chips for inference tasks without additional re-training steps.

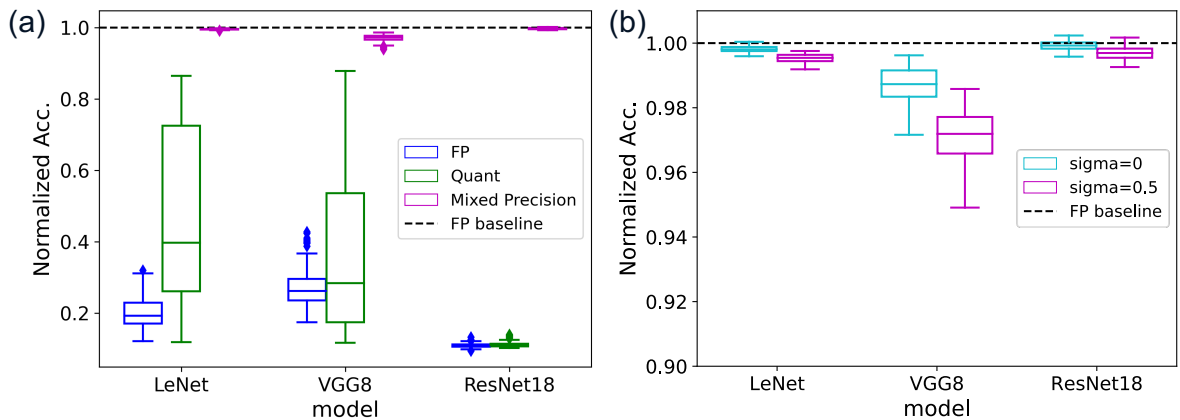

**Figure 7.** (a) Boxplot showing the accuracy of the FP, quantization-aware and mixed-precision trained models after weight transfer to another CIM chip. The hardware parameters are included in Table 1. Programming error with sigma = 0.5 is injected during model transfer. Models trained with floating point precision and quantization-aware training suffer from significant accuracy loss after model transfer, whereas the on-chip trained models maintain software comparable accuracy. Each boxplot contains 100 trials. (b) Boxplot showing the accuracy of the mixed-precision trained models for different programming error levels during weight transfer.

## 2.7. Evaluation of System Energy Efficiency

In this section, we assess the energy efficiency of the CIM chip. Our complete system achieves peak 3.08 TOPS/W (98.6 TOPS/W if normalized to 1bit input and 1bit weight) per tile at the clock frequency of 100MHz. The energy consumed for one VMM operation per tile (Energy/Tile Op) is 2.66nJ. For MNIST on LeNet, the energy consumption during inference is estimated to be 1.9μJ/image and 0.46ms/image. Total energy and latency are assessed by adding the required energy and time to finish the VMM operations of each layer. It should be mentioned that some intermediate processing steps, like activation function, batch normalization, pooling function and scaling factor multiplication are performed in the digital unit and are not included in the above estimation. For VGG-8 and ResNet-18, we expand the array size to 256×64 and included the increased bit-line current associated with the larger array size to 2.93nJ per array operation. For the CIFAR-10 inference task, it takes 0.023mJ and 1.94ms to process one image on VGG-8, 0.24mJ and 4.9ms on ResNet-18.

The average latency per image can be greatly reduced if we adopt inter-layer pipelining, which begins the computation of one layer as soon as sufficient activations from the previous layer have been computed. The average latency per image can be greatly reduced to 0.42ms for LeNet, 0.74ms for VGG-8 and 0.75ms for ResNet-18. With pipelining, the latency is determined by the slowest layer. By creating weight copies for slow layers (i.e. the first few convolution layers), we can balance the computation throughput of the slow layers and fast layers, therefore further reducing the average latency. For LeNet, the latency of Conv1 layer is 0.415ms, whereas the rest two layers only take 0.046ms and 0.001ms. To match the latency, we create 8 copies for Conv1 layer. Since the weight matrix of Conv1 is small (25×8), two copies can share the same tile, adding only 3 extra crossbar arrays. The inference can be accelerated by 7.9× from 0.42ms to 0.053ms. Similarly, for VGG-8, 4 parallel copies of the first two convolution layers are created, achieving a speedup of 3.9× with an overhead of 7 additional arrays. For ResNet-18, the latency of first five layers is 0.737ms. To match the latency of faster layers (0.184ms), the first five layers are copied 4 times and mapped to 74 additional crossbar arrays. The average latency is reduced from 0.75ms to 0.2ms.

The energy consumption of the CIM SoC is dominated by the mixed-signal circuit, where the crossbar arrays only account for a small fraction of the total energy (< 2%[5d]). The energy consumption can be reduced by about 6× with optimized mixed-signal circuit. We recently reported a more optimized circuit design[5d], which is fabricated in the same process and shares similar architecture. The optimized design achieves 20.7 peak TOPS/W (662.4 normalized TOPS/W), suggesting better energy efficiency can be achieved with the proposed hybrid training method.

**Table 2.** Time and energy consumption estimation of MAC inference on analog CIM.

| Model | # of Devices | FLOPs | # of Crossbars | # of Crossbars with Copies | # of Ops | # of Tile Ops | Latency (ms) | Latency with Pipelining (ms) | Latency with Copies (ms) | Energy per Image (mJ) |
|---|---|---|---|---|---|---|---|---|---|---|
| LeNet | 6.4k | 273k | 6 | 9 | 641 | 707 | 0.46 | 0.42 | 0.05 | 0.0019 |
| VGG-8 | 1.1M | 77.8M | 78 | 85 | 2690 | 7713 | 1.94 | 0.74 | 0.19 | 0.023 |
| ResNet-18 | 22.3M | 1.1G | 1480 | 1554 | 6801 | 81922 | 4.90 | 0.75 | 0.20 | 0.24 |

## 3. Discussion and Conclusion

In this work, we develop a mixed-precision analog/digital system for DNN training. A bulk-type RRAM device that exhibits features such as forming-free, gradual switching, low noise and low operation currents, is used for VMM operation in the forward pass. Weight gradient is calculated and accumulated in a digital unit. The RRAM conductance is modified only when the accumulated gradient crosses a defined update threshold. This method maintains the gradient accumulation in high precision and enables sparse weight updates. The gradient accumulation significantly relaxes the device requirements, including conductance update granularity, linearity, symmetry and endurance, and supports a variety of optimizers.

A fully integrated SoC with multiple CIM tiles has been developed in 65nm technology that supports bitcell programming and neural network applications. Device and system-level variabilities are characterized and modelled in a hardware-aware simulator developed in the PyTorch framework. To achieve high throughput and low latency, the circuit is designed to operate in the bit-serial fashion. The complete system achieves 3.08TOPS/W (98.6TOPS/W if normalized to 1bit input and 1bit weight) per tile at the clock frequency of 100MHz.

We experimentally demonstrate effective model training using the mixed-precision setup. With only 2-bit weight precision during CIM inference, the LeNet model is able to converge in 13 epochs, and the number of weight updates is reduced by 500×. The proposed system for training of larger neural networks was further evaluated in a hardware-aware simulator which considers hardware limitations of all levels. VGG-8 and ResNet-18 trained in the mixed precision approach can achieve software-comparable accuracy with only 4-bit weights and less than 1% weight updates compared to those trained in software. Additionally, the trained network is robust to programming errors during weight transfer, therefore eliminating the need of network re-training.

## 4. Experimental Section

*b-RRAM Device Fabrication*: The bilayer bulk RRAM stack includes a wide-band dielectric layer and a metal oxide layer, which are deposited by ALD and PVD in sequence without air break using AMAT's deposition platform. The diameter of the bitcells that are integrated on top of the CMOS peripheral circuits is 200nm.

*Device Characterization*: A custom PCB board was designed for device testing, silicon bring-up and DNN training demonstrations. Additional on-board DACs and ADCs provides the current and voltage references and extra testing capabilities. To store the large input dataset, an SPI flash is also added on board, which sends the input images to the data memory through SPI. A Raspberry Pi is connected to the PCB via SPI that initiates commands to on-chip processor and collects VMM outputs from the chip.

*Bit-serial DAC and ADC*: To improve the throughput while eliminating conductance non-linear effects, we adopt the bit-serial operation. The input is converted to an 8-bit serial pulse train with a fixed read voltage amplitude. The $n^{th}$ input pulse is applied to the Drive Line at the $n^{th}$ cycle according to the $n^{th}$ bit of the quantized input. No input voltage will be applied for "0" input digit, and the defined read voltage pulse will be applied for "1" input digit. The output is converted by the TIA and ADC. At each cycle, the TIA first converts the bit-line current to voltage, then the binary-weighted multi-cycle sampling ADC samples the voltage and adds it the previous sampled value. The accumulated value will then be halved. After 8 sampling cycles, a single conversion step will be applied to convert the weighted sum to a single 8-bit output value. The overall ADC behavior can be depicted with the following equation:

$$D_{OUT} = 2^{-1}V_{IN}[8] + 2^{-2}V_{IN}[7] + \cdots + 2^{-8}V_{IN}[1] \qquad (1)$$

*Hardware-aware Simulation*: To support mixed-precision training and evaluate the network robustness to weight perturbation, a hardware-aware simulation platform is developed in the PyTorch framework. Figure S3 shows the hierarchical organization of the simulator. The simulator takes the neural network architecture and hardware models as inputs and then automatically convert the Conv/FC modules to the corresponding custom Conv/FC modules that can be deployed on the CIM hardware.

The hardware constraints are modeled in three levels. Level-1 introduces quantization, weight noise and limited on/off ratio. A fake-quantization function will quantize the FP32 number to discrete values of FP32 number. The gradient is computed using the empirical straight through estimator[24]. As the name suggested, it ignores the non-differentiable threshold function and passes on the incoming gradient as an identity function. When performing noise-aware training, noise is randomly sampled from a defined Gaussian distribution and added to the weights for the convolution. In Level-2, differential column mapping included. The weight matrix is split into the positive and negative part, takes the absolute value and adds the minimum weight. The output computed with the negative weight will be subtracted from that of the positive weight. In Level-3, we consider the limited array size and slice convolution kernels that are too large to multiple arrays. The outputs of each tile are quantized by ADC and added together to get the final output. Here, we assume ADC has fixed input current range.

For mixed precision training, a weight change matrix $\Delta W_{FP}$ is used to accumulate the weight update. The corresponding weight update will be transferred to the b-RRAM device conductance if it crosses a predefined threshold after each batch. The element in $\Delta W_{FP}$ is then reset to zero. To simulate programming error during RRAM weight update, we inject error by sampling from a Gaussian distribution obtained from on-chip measurements.

*Energy Efficiency*: The energy efficiency of the SoC is evaluated by TOPS/W. When operated at 100MHz, the energy consumption to perform the VMM operations on the entire tile is 2.66nJ. The peak TOPS/W at 100% array utilization is computed as 3.08 TOPS/W with the following equation:

$$\frac{\text{TOPS}}{\text{W}} = \frac{Number\ of\ Operations/Sec}{Power\ Consumption} = \frac{2\times Number\ of\ MAC/Sec}{Power\ Consumption} \qquad (2)$$

The energy consumption to perform inference on one image is calculated by multiplying 2.66nJ with the total number of Tile Ops in Table 2. Here, a Tile Op means the VMM operations on one entire tile.

$$\frac{Power\ Consumption}{Image} = \text{Power Compution/Tile Op} \times Number of\ Tile\ Ops \qquad (3)$$

It should be noted that the intermediate processing steps, like activation function, batch normalization, pooling function and scaling factor multiplication are computed in the digital unit and are not included in the above energy consumption calculation. These computation blocks can be supported by the on-chip processor in the future and accounted for in the energy efficiency consumption. The system performance can be further improved by optimizing the peripheral circuits and fabricating with more advanced technology nodes. For example, an optimized CIM SoC with same technology node and same functionalities achieves 20.7 peak TOPS/W[5d].

**Acknowledgements**

The authors would like to thank Dr. Fuxi Cai, Dr. Justin M. Correll, Dr. Deepak Kamalanathan and Joe Hsu for helpful discussions. This work was supported in part by Applied Materials and by the National Science Foundation through Award # CCF-1900675.

Received: ((will be filled in by the editorial staff))
Revised: ((will be filled in by the editorial staff))
Published online: ((will be filled in by the editorial staff))